\documentclass[aps,prl,amsmath,amssymb,twocolumn,superscriptaddress]{revtex4}
\usepackage{graphicx}
\usepackage{braket}
\usepackage{color}
\usepackage{hyperref}
\usepackage{soul}

\begin{document}


\title{Symmetric Operation of the Resonant Exchange Qubit}

\author{Filip~K.~Malinowski}
\thanks{These authors contributed equally to this work}
\affiliation{Center for Quantum Devices, Niels Bohr Institute, University of Copenhagen, 2100 Copenhagen, Denmark}

\author{Frederico~Martins}
\thanks{These authors contributed equally to this work}
\affiliation{Center for Quantum Devices, Niels Bohr Institute, University of Copenhagen, 2100 Copenhagen, Denmark}

\author{Peter~D.~Nissen}
\affiliation{Center for Quantum Devices, Niels Bohr Institute, University of Copenhagen, 2100 Copenhagen, Denmark}

\author{Saeed~Fallahi}
\affiliation{Department of Physics and Astronomy, Birck Nanotechnology Center, Purdue University, West Lafayette, Indiana 47907, USA}

\author{Geoffrey~C.~Gardner}
\affiliation{Department of Physics and Astronomy, Birck Nanotechnology Center, Purdue University, West Lafayette, Indiana 47907, USA}
\affiliation{School of Materials Engineering and School of Electrical and Computer Engineering, Purdue University, West Lafayette, Indiana 47907, USA}

\author{Michael~J.~Manfra}
\affiliation{Department of Physics and Astronomy, Birck Nanotechnology Center, and Station Q Purdue, Purdue University, West Lafayette, Indiana 47907, USA}
\affiliation{School of Materials Engineering, Purdue University, West Lafayette, Indiana 47907, USA}

\author{Charles~M.~Marcus}
\affiliation{Center for Quantum Devices and Station Q Copenhagen, Niels Bohr Institute, University of Copenhagen, 2100 Copenhagen, Denmark}

\author{Ferdinand~Kuemmeth}
\affiliation{Center for Quantum Devices, Niels Bohr Institute, University of Copenhagen, 2100 Copenhagen, Denmark}

\newcommand{\VLP}{V_\mathrm{LP}}
\newcommand{\VLB}{V_\mathrm{LB}}
\newcommand{\VMP}{V_\mathrm{MP}}
\newcommand{\VRB}{V_\mathrm{RB}}
\newcommand{\VRP}{V_\mathrm{RP}}

\newcommand{\uud}{\uparrow\uparrow\downarrow}
\newcommand{\udu}{\uparrow\downarrow\uparrow}
\newcommand{\duu}{\downarrow\uparrow\uparrow}

\newcommand{\TR}{T_\mathrm{R}}
\newcommand{\fR}{f_\mathrm{R}}
\newcommand{\TCPMG}{T_2 ^\mathrm{CPMG}}
\renewcommand{\vec}[1]{{\bf #1}}

\date{\today}

\begin{abstract}

We operate a resonant exchange qubit in a highly symmetric triple-dot configuration using IQ-modulated RF pulses. 
At the resulting three-dimensional sweet spot the qubit splitting is an order of magnitude less sensitive to all relevant control voltages, compared to the conventional operating point, but we observe no significant improvement in the quality of Rabi oscillations. 
For weak driving this is consistent with Overhauser field fluctuations modulating the qubit splitting. 
For strong driving we infer that effective voltage noise modulates the coupling strength between RF drive and the qubit, thereby quickening Rabi decay. 
Application of CPMG dynamical decoupling sequences consisting of up to $n=32$ $\pi$ pulses significantly prolongs qubit coherence, leading to marginally longer dephasing times in the symmetric configuration. 
This is consistent with dynamical decoupling from low frequency noise, but quantitatively cannot be explained by effective gate voltage noise and Overhauser field fluctuations alone.  
Our results inform recent strategies for the utilization of partial sweet spots in the operation and long-distance coupling of triple-dot qubits.

\end{abstract}

\maketitle

Spin qubits are widely investigated for applications in quantum computation~\cite{Loss1998,Petta2005,Shulman2012,Koppens2006,Nowack2011,Veldhorst2015,Nichol2017}, with several operational choices depending on whether the qubit is encoded in the spin state of one~\cite{Koppens2006,Nowack2011,Veldhorst2015,Kawakami2016,Maurand2016,Takeda2016}, two~\cite{Petta2005,Shulman2012,Foletti2010,Nichol2017} or three electrons~\cite{Laird2010,Gaudreau2011,Medford2013,Medford2013a,Kim2014,Eng2015,Russ2016}. 
In particular, spin qubits encoded in three-electron triple quantum dots allow universal electrical control with voltage pulses, and enable integration with superconducting cavities~\cite{Petersson2012,Stockklauser2015,Liu2015,Russ2015a,Srinivasa2016,Mi2016}. 
Multi-qubit coupling via superconducting cavities, however, is challenging due to the effects of environmental noise on resonant exchange (RX) qubits \cite{Medford2013a,Srinivasa2016}.
A recent approach to improve coherence times is the operation at sweet spots, where the qubit splitting is to first order insensitive to most noisy parameters~\cite{Martins2016,Reed2016,Russ2015,Shim2016a}. 
Here, we operate a symmetric resonant exchange (SRX) qubit in which the qubit splitting is highly insensitive to all three single-particle energies~\cite{Shim2016a}, and compare its performance to its conventional configuration as a RX qubit~\cite{Medford2013a,Taylor2013}.

\begin{figure}
	\includegraphics[width=0.5\textwidth]{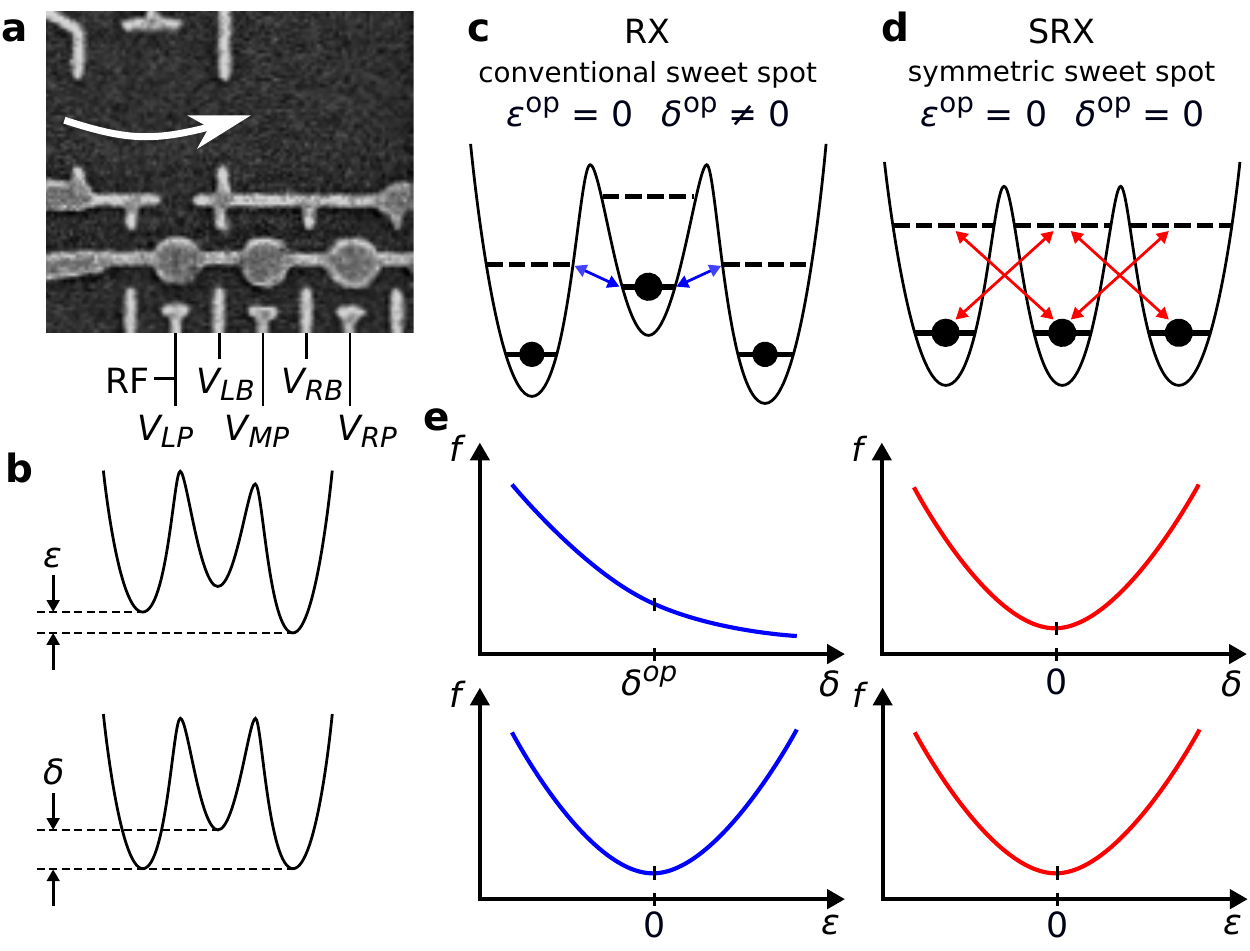}
	\caption{(a) Scanning electron micrograph of a GaAs triple quantum dot, formed under the rounded accumulation gate, and a proximal sensor dot (white arrow), formed by depletion gates. The five depletion gates used for qubit manipulation are labeled.  
(b) Schematic illustration of two control parameters, $\delta$ and $\varepsilon$, resulting in energy shifts $\delta|e|$ and $\varepsilon|e|$.
(c) Potential along the RX qubit. The qubit splitting arises from virtual tunneling of the central electron to the outer dots (blue arrows), and is therefore sensitive to potential fluctuations of each dot. 
(d) Potential along the SRX qubit. Tunneling of the outer electrons to the central dot contributes to charge hybridization equally strongly as tunneling of the central electron to the outer dots (red arrows), making the qubit splitting insensitive to potential fluctuations of all three dots. 
(e) Schematic dependence of the qubit frequency $f$ on $\varepsilon$ and $\delta$ around the operating point of the RX and SRX qubit.
	}
	\label{fig1}
\end{figure}

We configure a triple-quantum-dot device either as a SRX or RX qubit by appropriate choice of gate voltages.
Gate electrodes are fabricated on a doped, high-mobility GaAs/AlGaAs quantum well,
and the triple dot is located $\sim 70$ nm below three circular portions of the accumulation gate (Fig.~\ref{fig1}a). 
The occupation of the dots is controlled on nanosecond timescales by voltage pulses on gates $V_i$, where $i$ refers to the left/middle/right plunger gate (LP/MP/RP) or left/right barrier gate (LB/RB). 
Radio frequency (RF) bursts for resonant qubit control are applied to the left plunger gate. 
The conductance through the proximal sensor dot is sensitive to the charge occupation of the triple quantum dot, allowing qubit readout (see below). 

In the presence of an in-plane magnetic field, $B=400$~mT in this experiment, the triple-dot qubit is defined by the two three-electron spin states with total spin $S=1/2$ and spin projection $S_z=1/2$
~\cite{Laird2010,Medford2013a,Taylor2013,Russ2016}.
Ignoring normalization, these spin states can be represented by $\ket{0} \propto (\ket{\duu}-\ket{\udu})+(\ket{\uud}-\ket{\udu})$ and $\ket{1} \propto (\ket{\uud}-\ket{\duu})$. 
Here, arrows indicate the spin of the electron located in the left, middle and right quantum dot.
Note that the spin state of  $\ket{0}$ and $\ket{1}$ is, respectively, symmetric and antisymmetric under exchange of the outer two electrons.
In the presence of interdot tunneling this exchange symmetry affects hybridization of the associated orbital wavefunctions, splitting $\ket{0}$ and $\ket{1}$ by $h f$ (where $h$ is Planck's constant and $f$ sets the frequency of the qubit's rotating frame).
Similarily, an additional triple-dot state with $S=3/2$ and $S_z = 1/2$ is split from the qubit states due to interdot tunneling. 
All other triple-dot states have different $S_z$ and are energetically separated from the qubit states due to the Zeeman effect. 

In the conventional operating regime of the RX qubit (Fig.~\ref{fig1}c) the (111) charge state of the triple dot is hybridized weakly with charge states (201) and (102) (here number triplets denote the charge occupancy of the triple dot). 
This lowers the energy of $\ket{0}$ with respect to $\ket{1}$ and makes the resulting qubit splitting sensitive to detuning of the central dot, $\delta$ (cf. Fig.~\ref{fig1}b,e)~\cite{Medford2013a}. The qubit splitting is, however, to first order insensitive to detuning between the outer dots, $\varepsilon$, \cite{Taylor2013}, reflecting that tunneling across left and right barrier contribute equally to the qubit splitting (Fig.~\ref{fig1}c,e).  
Qubit rotations in the rotating frame are implemented by applying RF bursts to gate $\VLP$, such that the operating point oscillates around $\varepsilon=0$. When the RF frequency matches the qubit splitting, the qubit nutates between $\ket{0}$ and $\ket{1}$, allowing universal control using IQ modulation \cite{Medford2013a}. 
When the detuning of the outer dots is ramped towards (201), $\ket{0}$ maps to a singlet state of the left pair ($\ket{S_L}\propto (\ket{\duu}-\ket{\udu})$, see first terms in $\ket{0}$), whereas $\ket{1}$ remains in the (111) charge state due to the Pauli exclusion principle \cite{Laird2010,Medford2013,Medford2013a}. 
This spin-to-charge conversion allows us to perform single-shot readout on microsecond timescales, by monitoring a proximal sensor dot using high-bandwidth reflectometry~\cite{Barthel2010a}. In this work we estimate the fraction of singlet outcomes, $P_\mathrm{S}$, by averaging 1000-10000 single-shot readouts.

In the case of the SRX qubit, however, all three single-particle levels are aligned, and the (111) state hybridizes with the charge states (201), (102), (120) and (021)~\cite{Shim2016a}. This introduces additional symmetries between the tunneling of the electron from the central dot to the outer dots and tunneling of the outer electrons to the central dot (Fig.~\ref{fig1}d). 
As a consequence, the qubit splitting is expected to be insensitive to first order to both $\varepsilon$ and $\delta$ (Fig.~\ref{fig1}e) as well as to the barrier detuning, $\varepsilon_B$ (introduced below).
Due to the required alignment of single-particle levels, hybridization is suppressed by the charging energy within each dot (indicated by the large energy spacing between solid and dashed lines in Fig.~\ref{fig1}c,d).
Accordingly, we find that much larger tunnel couplings have to be tuned up to maintain a significant qubit splitting. 
In practice, the gate voltage configuration needed to achieve a SRX qubit splitting of a few hundred megahertz does no longer allow spin-to-charge conversion solely by a ramp of $\varepsilon$. Therefore, we also apply voltage pulses to the barrier gates when ramping the qubit between the operation configuration (indicated by superscript $\mathrm{op}$) and readout configuration (see below).

\begin{figure}
	\includegraphics[width=0.5\textwidth]{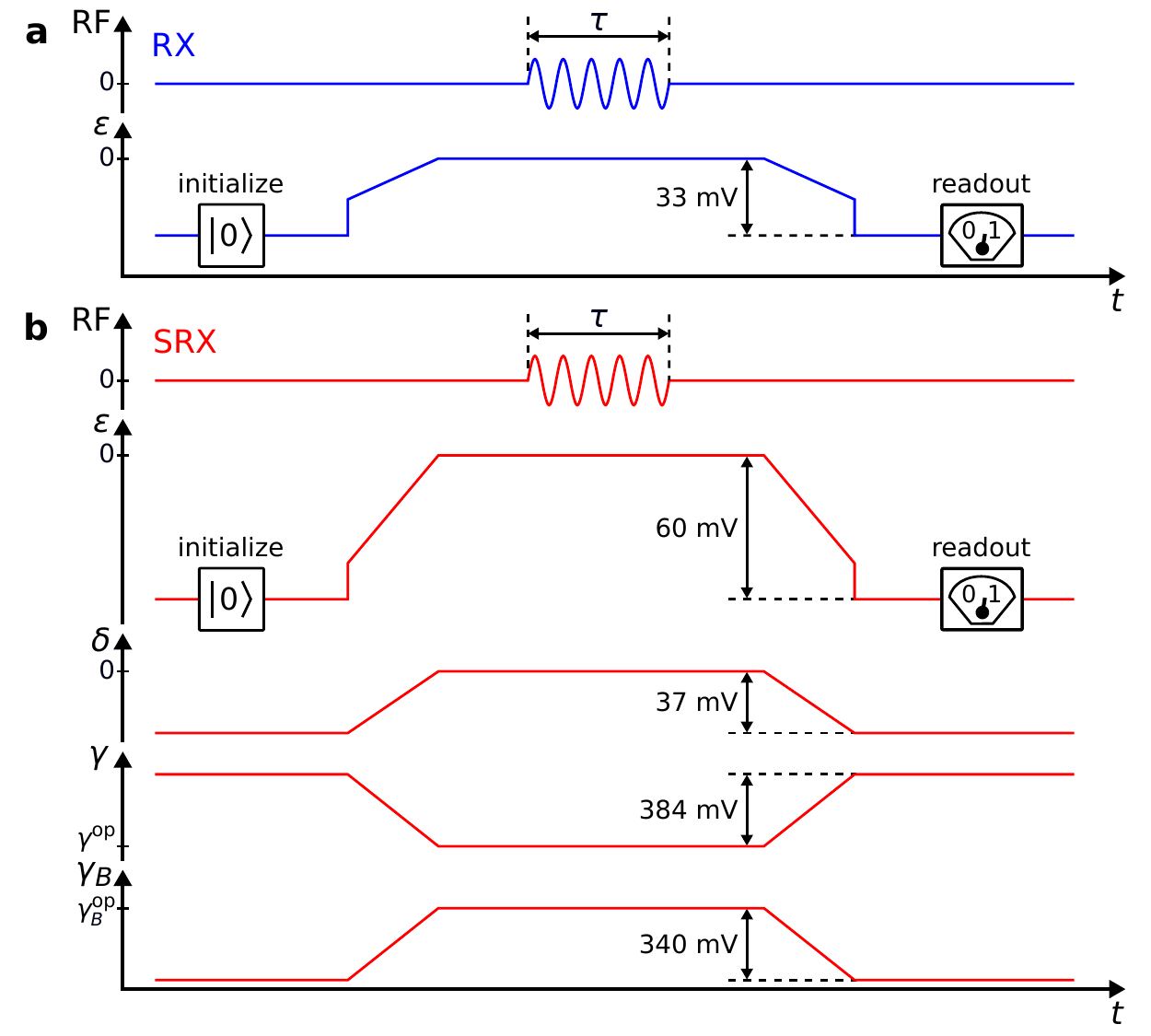}
	\caption{
Schematic pulse cycle for measuring Rabi oscillations of the RX (a) and SRX (b) qubit. 
An IQ-modulated RF burst is applied on resonance with the qubit splitting for duration $\tau$. 
Linear detuning ramps, with typical amplitudes indicated, implement spin-to-charge conversion needed for qubit initialization and readout. 
For qubit spectroscopy and CMPG measurements the RF burst is replaced by a continuous RF tone or a sequence of calibrated RF pulses, respectively.
	}
	\label{fig2}
\end{figure}

Figure ~\ref{fig2}a (\ref{fig2}b) defines the pulse cycle used for spectroscopy and operation of the RX (SRX) qubit.
Taking into account the physical symmetries of the device (cf. Fig.~\ref{fig1}), control parameters $ \varepsilon, \gamma, \delta, \varepsilon_B, \gamma_B $ are specified in terms of gate voltages $V_i$,

\begin{equation}
	\left(
	\begin{array}{c}
		\varepsilon \\
		\delta \\
		\gamma
	\end{array}
	\right) = \frac{1}{\sqrt{6}}\left(
	\begin{array}{ccccc}
		-\sqrt{3} & 0 & \sqrt{3} \\
		-1 & 2 & -1 \\
		\sqrt{2} & \sqrt{2} & \sqrt{2}
	\end{array}
	\right) \left(
	\begin{array}{c}
		\VLP-\VLP^\mathrm{sym} \\
		\VMP-\VMP^\mathrm{sym} \\
		\VRP-\VRP^\mathrm{sym}
	\end{array}
	\right)
	\nonumber
	\label{parameters1}
\end{equation}
\begin{equation}
	\left(
	\begin{array}{c}
		\varepsilon_B \\
		\gamma_B
	\end{array}
	\right) = \frac{1}{\sqrt{6}}\left(
	\begin{array}{ccccc}
		-\sqrt{3} & \sqrt{3} \\
		\sqrt{3} & \sqrt{3}
	\end{array}
	\right) \left(
	\begin{array}{c}
		\VLB-\VLB^\mathrm{sym} \\
		\VRB-\VRB^\mathrm{sym}
	\end{array}
	\right),
	\nonumber
	\label{parameters2}
\end{equation}
and the power ($P_\mathrm{RF}$), duration ($\tau$), frequency ($f_\mathrm{RF}$) and phase of the IQ-modulated RF burst. 
The operating point of the SRX qubit, defined by $V_i=V_i^\mathrm{sym}$, was chosen to yield a qubit frequency of 530~MHz
~\footnote{($ \VLP^\mathrm{sym}, \VLB^\mathrm{sym}, \VMP^\mathrm{sym}, \VRB^\mathrm{sym}, \VRP^\mathrm{sym} $) = (-1.53, -0.14, -1.05, -0.15, -0.83) V.}. 
The operating point of the RX qubit, located at  $\{ \delta^\mathrm{op}>0, \gamma^\mathrm{op}>0, \gamma_B^\mathrm{op}<0\}$, was chosen to yield a comparable qubit frequency of 510~MHz.
The linear ramps before (after) the RF burst facilitate initialization (readout) of the qubit state via an adiabatic conversion of a two-electron spin singlet state in the left dot.
For the RX qubit $\{ \delta-\delta^\mathrm{op}, \gamma-\gamma^\mathrm{op}, \varepsilon_B, \gamma_B-\gamma_B^\mathrm{op} \}$ all remain zero throughout the pulse cycle, i.e. the operation and readout configuration differ only in detuning $\varepsilon$ (Fig.~\ref{fig2}a). In contrast, to adiabiatically connect the initialization/readout point of the SRX qubit to its operating point, we found it necessary to vary $\varepsilon$, $\delta$, $\gamma$ and $\gamma_B$ during the pulse cycle (Fig.~\ref{fig2}b), which involves voltage pulses on all five gates indicated in Fig.~\ref{fig1}a.

\begin{figure}
	\includegraphics[width=0.5\textwidth]{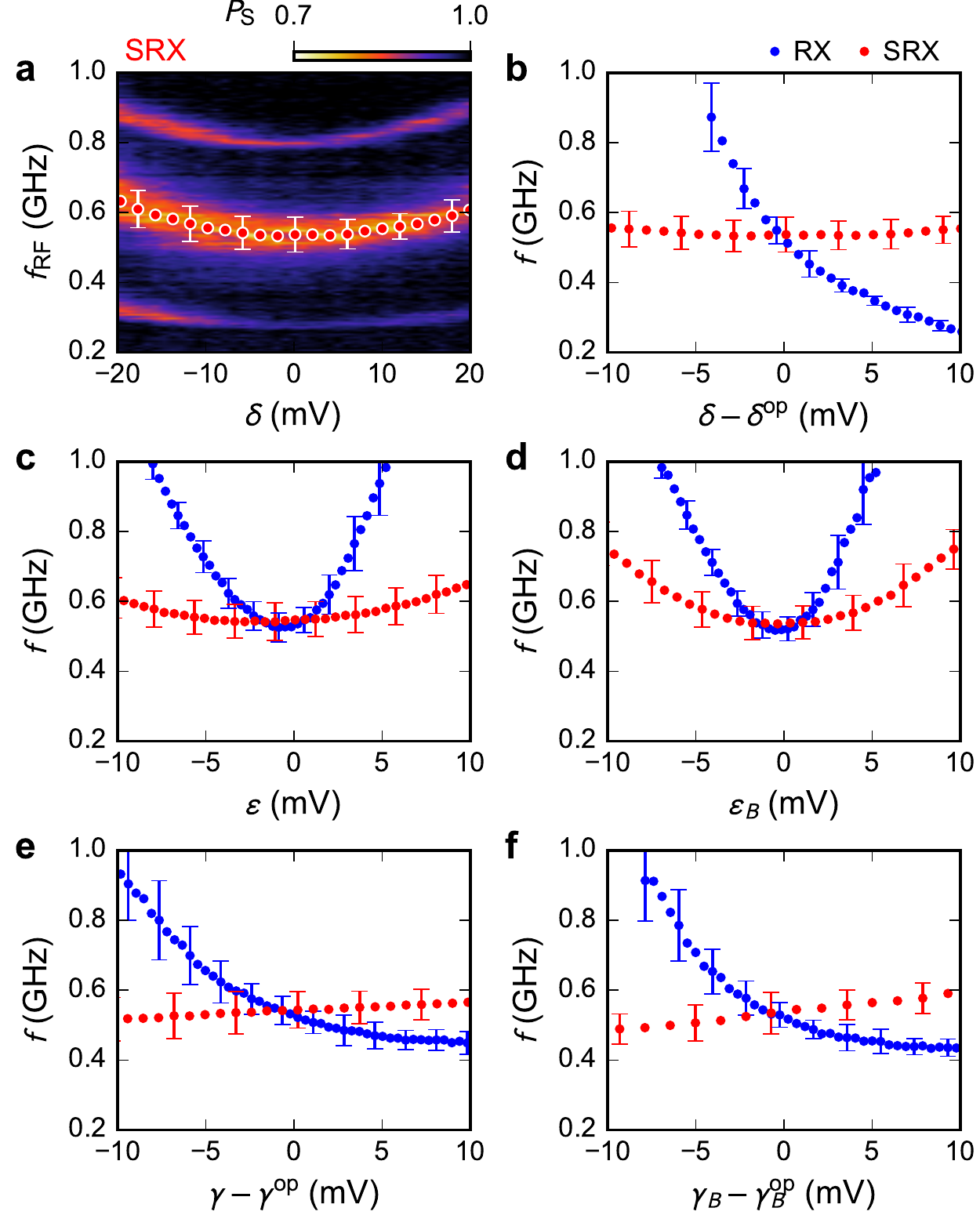}
	\caption{(a) Qubit spectroscopy along $\delta$ around the SRX operating point (see text). The red circles indicate the extracted qubit splitting $f$. Additional resonances correspond to multiphoton excitations of the triple dot. (b-f) Extracted qubit splitting along $\delta$, $\varepsilon$, $\varepsilon_B$, $\gamma$ and $\gamma_B$ for the SRX (red) and RX (blue) configuration, around their corresponding operating points. 
	Error bars indicate the inhomogeneous line width of the resonance.
	}
	\label{fig3}
\end{figure}

Qubit spectroscopy performed in the vicinity of the operating point quantitatively reveals each qubit's symmetries and susceptibilities to gate voltage fluctuations. First, maps as in Fig~\ref{fig3}a are acquired by repeating a pulse cycle with $\tau=150$~ns fixed, and plotting the fraction of singlet readouts, $P_\mathrm{S}$, as a function of $f_\mathrm{RF}$, while stepping the control parameters along five orthogonal axes that intersect with the operating point. 
The qubit frequency $f$ is extracted from the center of the dominant $P_\mathrm{S}(f_\mathrm{RF})$ resonance (cf. red circles in Fig~\ref{fig3}a), and plotted as a function of $\varepsilon$, $\delta$, $\gamma$, $\varepsilon_B$ and $\gamma_B$ (Fig.~\ref{fig3}b-f).
Indeed, the dependence of $f$ on $\delta$ reveals that the SRX qubit splitting is to first order insensitive to $\delta$, in contrast to the conventional RX qubit (Fig.~\ref{fig3}b). Further, we observe that both qubits show a sweet spot with respect to $\varepsilon$ and $\varepsilon_B$ (Fig.~\ref{fig3}c,d), indicating that the symmetry breaking associated with $\varepsilon_B\neq0$ is analogues to the well-known symmetry breaking associated with $\varepsilon\neq0$~\cite{Taylor2013}. 
Interestingly, for both detuning parameters, the curvature of the qubit splitting is significantly smaller for the SRX configuration, compared to the RX configuration. 
Moreover, the SRX qubit frequency is also significantly less susceptible to changes in parameters $\gamma$ and $\gamma_B$, compared to the conventional RX qubit (Fig.~\ref{fig3}e,f), corroborating the potential use of this highly symmetric configuration for prolonging qubit coherence. 

The qubit spectra from Figures \ref{fig3} allow us to quantify the susceptibility of the qubit splitting to gate voltage fluctuations, by evaluating
\begin{equation}
	S = \sqrt{ \sum\limits_{\begin{subarray}{c}
	i \in \{\mathrm{LP, LB,} \\ \mathrm{MP, RB, RP}\}
	\end{subarray}}
	 \left( \frac{\partial f}{\partial V_i} \right)^2}
	 = \sqrt{ \sum\limits_{\begin{subarray}{c}
	\xi \in \{\varepsilon, \delta, \\ \gamma, \varepsilon_B, \gamma_B \}
	\end{subarray}}
	 \left( \frac{\partial f}{\partial \xi} \right)^2
	 }
\end{equation}
for both operating points. 
For the SRX qubit we find a susceptibility to gate noise ($S = 6$~MHz/mV) that is one order of magnitude smaller compared to the RX qubit ($S = 66$~MHz/mV). 
For the linear coupling regime this means that voltage fluctuations on gate electrodes, including instrumentation noise propagating on the cryostats wideband transmission lines, are expected to be much less detrimental to the SRX qubit than to the RX qubit.

\begin{figure}
	\includegraphics[width=0.5\textwidth]{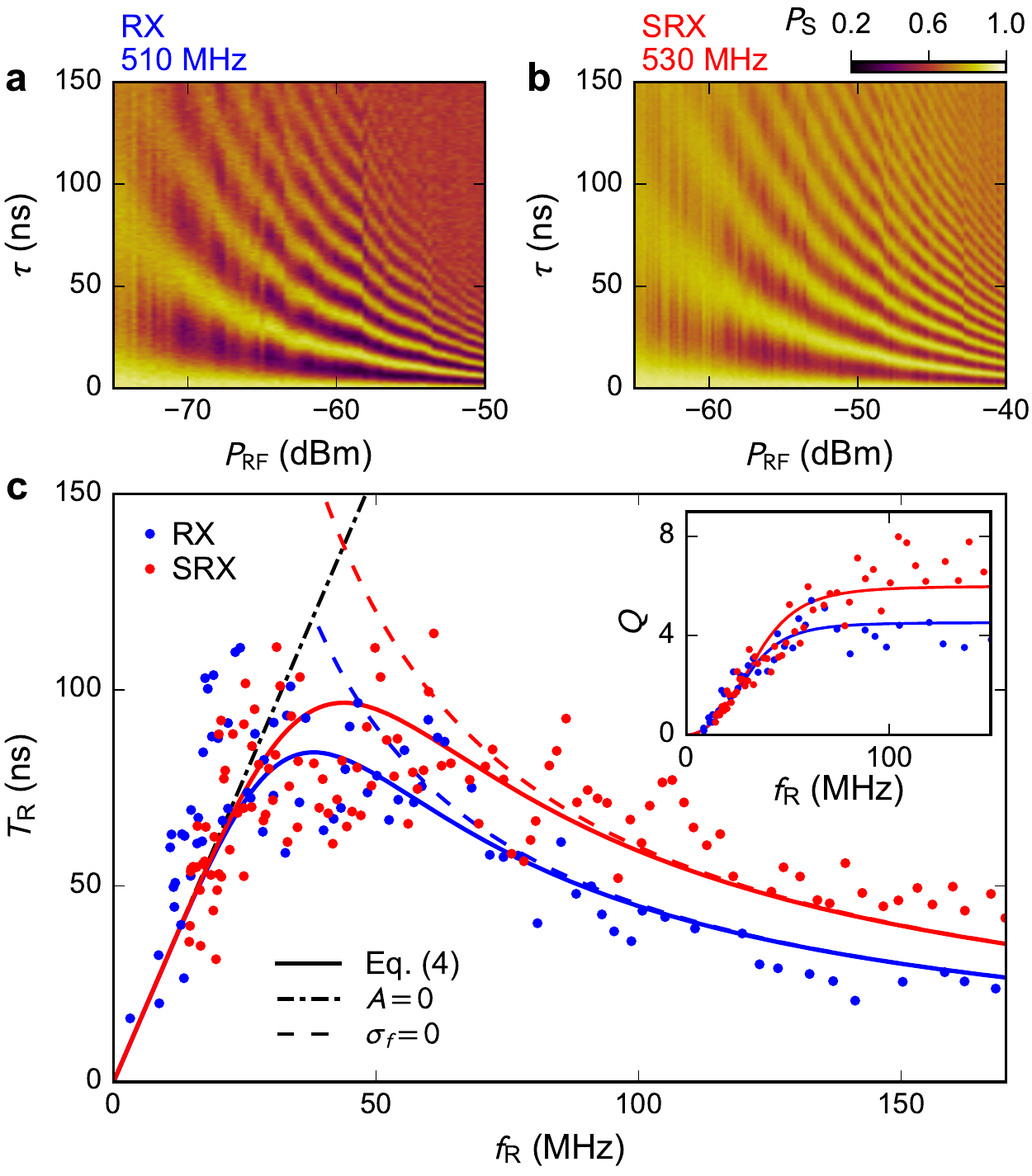}
	\caption{(a,b) Rabi oscillations of the RX and SRX qubit as a function of RF burst time ($\tau$) and excitation power ($P_\mathrm{RF}$) obtained at nearly identical qubit splitting of 510 MHz (RX) and 530 MHz (SRX).
	(c) Parametric plot of Rabi decay time $\TR$ and quality factor $Q$ (inset) as a function of Rabi frequency $\fR$, extracted from vertical cuts of (a) and (b). Solid lines are theory fits based on Eq.~\ref{eqTR} and $Q\equiv\TR \times \fR$. Broken lines indicate the limits imposed by solely detuning noise (black) or solely drive noise (red and blue).
	}
	\label{fig4}
\end{figure}

Next we investigate whether the reduced noise susceptibility of the SRX qubit results in improved Rabi oscillations (Fig.~\ref{fig4}). 
To achieve a comparable Rabi frequency, $\fR$,  we find that $P_\mathrm{RF}$ needs to be 10 dB larger for the SRX qubit compared to the RX qubit, consistent with the smaller curvatures observed in Fig.~\ref{fig3}. However, only for high $P_\mathrm{RF}$ do we observe improvement in SRX qubit performance relative to the RX qubit. For quantitative comparison we fit an exponentially damped cosine to $P_\mathrm{S}(\tau)$ for each RF power. Figure~\ref{fig4}c parametrically plots the extracted $1/e$ decay time ($\TR$) and quality factor ($Q=\TR \times \fR$) of Rabi oscillations as a function of $\fR$. For $\fR<50$~MHz the quality of SRX Rabi oscillations is comparable to the RX qubit, while for $\fR>50$~MHz $\TR$ and $Q$ are enhanced by approximately 50\%, relative to the RX qubit.

The marginal performance improvement observed for the SRX qubit can be analyzed quantitatively by extending theory from Ref.~\onlinecite{Taylor2013} to include the dependence of the Rabi oscillations decay time $\TR$ on the Rabi frequency $\fR$. Assuming quasistatic gate-voltage noise and quasistatic nuclear spin noise, we derive
\begin{equation}
	\label{eqTR}
	\left( \frac{1}{\TR} \right)^2 =  \frac{\sigma_f^4}{4 \fR^2} + \fR^2 A^2,
\end{equation}
where $\sigma_f$ quantifies the rms deviation of $f$ from $f_\mathrm{RF}$ due to effective voltage flucuations and Overhauser field fluctuations (discussed below). The quantity $A^2$ captures the effect of voltage fluctuations on the coupling strength of the RF drive
\begin{equation}
	\label{eqA2}
	A^2 = \frac{8 \pi}{\eta^2}
	\sum\limits_{\begin{subarray}{c}
	\xi = \varepsilon, \delta, \\ \gamma, \varepsilon_B, \gamma_B
	\end{subarray}}
	\left(
	\frac{\partial \eta}{\partial \xi} \sigma_\xi
	\right)^2,
\end{equation}
with $\sigma_\xi$ being the standard deviation of the fluctuating paramater $\xi$ and $\eta$ being the lever arm between amplitude of the RF drive and the qubit nutation speed in the rotating frame. 
We find that the observed $\TR (f_\mathrm{R})$ is well fitted by our theoretical model, using $A= 0.17$ (0.22) for the SRX (RX) qubit and a common value $\sigma_f= 0.025$ (solid lines in Fig.~\ref{fig4}c). Although Ref.~\onlinecite{Taylor2013} formally identified $\eta$ with 
\begin{equation}
	\label{eqeta}
	\eta = \sqrt{ \left( \frac{\partial J}{\partial V_{LP}} \right)^2 + 3 \left( \frac{\partial j}{\partial V_{LP}} \right)^2 }
\end{equation}
(here $J = (J_\mathrm{L }+ J_\mathrm{R})/2$ and $j = (J_\mathrm{L} - J_\mathrm{R})/2$ are symmetry-adapted exchange energies arising from exchange $J_\mathrm{L/R}$ between central and left/right dot), its implications for the properties of the $A^2$ term and associated Rabi coherence were not considered. 
Equations~(\ref{eqA2},\ref{eqeta}) would in principle allow the extraction of voltage noise in more detail, but experimentally the partial derivatives are not easily accessible. 
However, by plotting the expected limit of $\TR$ if only detuning noise (black dash-dotted line) or only drive noise (red and blue dashed lines) is modeled, we deduce that the dominating contribution to $\sigma_f$ arises not from effective gate voltage noise, but from fluctuations of the Overhauser gradient between dots.    
Assigning $\sigma_f= 0.025$ entirely to Overhauser fluctuations, we estimate the rms Overhauser field in each dot to be approximately 4.2~mT, in good agreement with previous work on GaAs triple dots~\cite{Medford2013a,Delbecq2016}.

The detrimental effect of fluctuating Overhauser fields on qubit dephasing is not surprising, given that the qubit states are encoded in the $S_z=1/2$ spin texture:
For $\ket{0}$ the spin angular momentum resides in the outer two dots, whereas for  $\ket{1}$ it resides in the central dot. This makes the qubit splitting to first order sensitive to Overhauser gradients between the central and outer dots \cite{Taylor2013}. Equation~\ref{eqeta} further suggests that the qubit drive strength depends on $j = (J_\mathrm{L} - J_\mathrm{R})/2$, which likely is first-order-sensitive to both $\varepsilon$ and $\varepsilon_B$, and hence we suspect that $f_\mathrm{R}$, unlike $f$, remains sensitive to the charge noise.
These conclusions suggest that triple-dot qubits will benefit from implementation in nuclear-spin-free semiconductors, and possibly from replacing IQ-control in the rotating frame by baseband voltage pulses. Recent theoretical work indicates that this may allow efficient two-qubit gates between neighboring qubits using exchange pulses~\cite{Shim2016a} and long-distance coupling via superconducting resonators~\cite{Russ2015a,Srinivasa2016}.

\begin{figure}
	\includegraphics[width=0.5\textwidth]{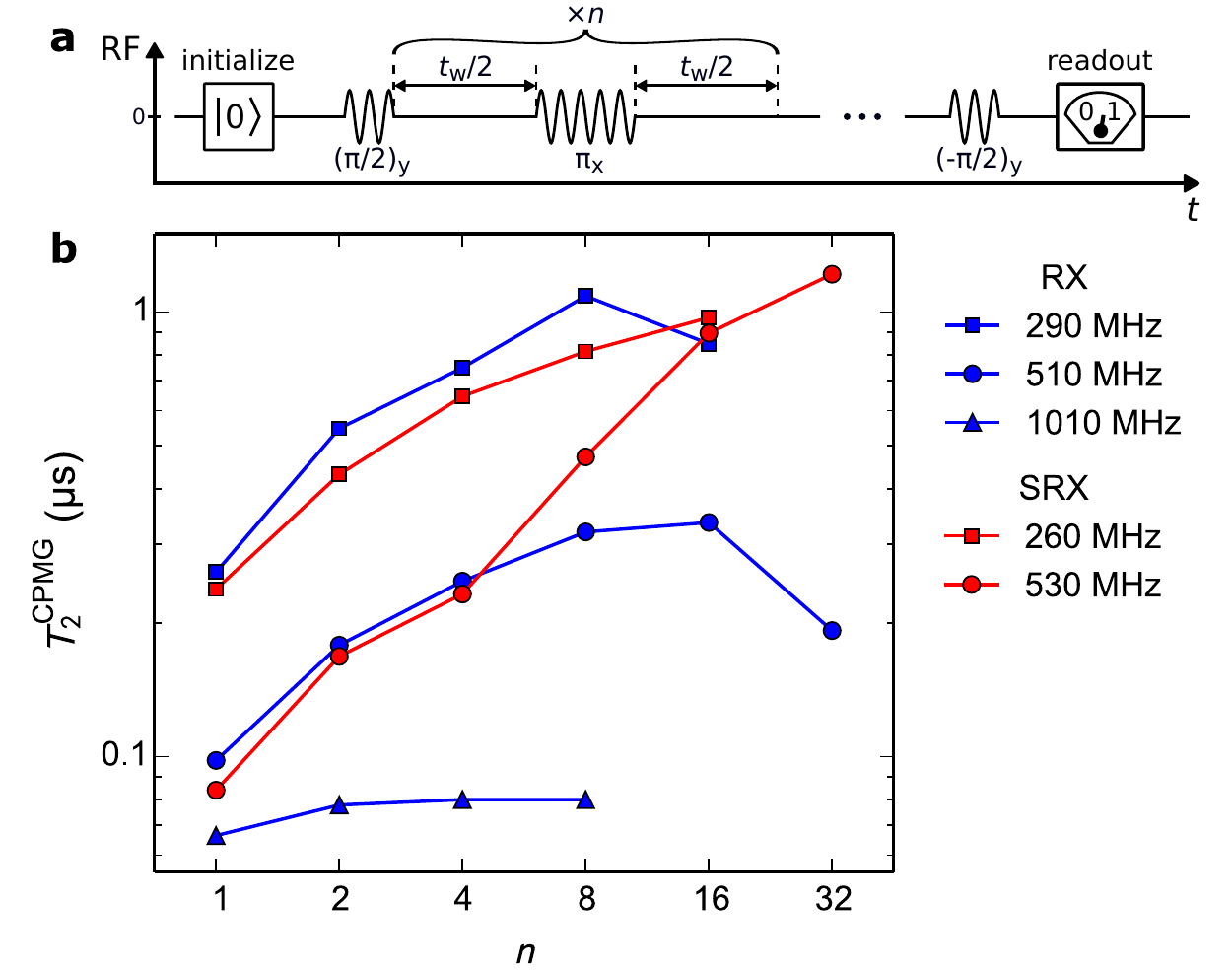}
	\caption{(a) CPMG dynamical decoupling sequence adapted from Ref.~\onlinecite{Medford2013a}. 
	The $(\pi/2)_y$ pulse prepares the superposition state $(1/\sqrt{2})(\ket{0}+\ket{1})$. The segment consisting of a waiting time, $t_W/2$, a $\pi_x$ pulse, and another waiting time, $t_W/2$, is repeated $n$ times ($n=1$ for Hahn echo). The $(-\pi/2)_y$ pulse projects the resulting state onto $\ket{0}$ or $\ket{1}$. The fraction of $\ket{0}$ outcomes, for increasing waiting time and fixed $n$, is used to extract the coherence time $T_2^\mathrm{CPMG}$ (see main text). 
	(b) $T_2^\mathrm{CPMG}$ as a function of the number of $\pi$ pulses for various  SRX and RX qubit frequencies.
	}
	\label{fig5}
\end{figure}

Finally, we test the prospect of the SRX qubit as a quantum memory, using Hahn echo and CPMG sequences consisting of relatively strong ($\tau \lesssim 10$ ns) $\pi$-pulses (defined in Fig.~\ref{fig5}a). 
These dynamical decoupling sequences are particularly effective against nuclear noise~\cite{Bluhm2011,Malinowski2017}, which is known to display relative long correlation times~\cite{Reilly2008,Barthel2009,Malinowski2017a}.
Figure \ref{fig5} shows the resulting coherence time, $T_2^\mathrm{CPMG}$, for different qubit frequencies, for up to $n$=32 pulses. Values for $T_2^\mathrm{CPMG}$ were extracted from Gaussian fits to $P_\mathrm{S}(T)$, where $T=n \cdot t_w$ is the total dephasing time. 
For small number of $\pi$-pulses we see no difference in the performance of the RX and SRX qubit, indicating that effective voltage noise (incl. instrumentation noise on gate electrodes) is not limiting coherence. Qualitatively, this may point towards high-frequency Overhauser fluctuations playing a dominant role, although we find coherence times significantly shorter than expected from nuclear spin noise alone~\cite{Petta2005,Bluhm2011,Malinowski2017a} and values reported for RX qubits \cite{Medford2013a}.  
While $\TCPMG$ strongly depends on the qubit frequency, the ratio $f \times \TCPMG$ is roughly independent of $f$ (not shown). This is reminiscent of gate defined quantum dots that showed a nearly exponential dependence of the exchange splitting on relevant control voltages ~\cite{Higginbotham2014,Veldhorst2015,Dial2013,Medford2013a,Martins2016,Reed2016}. 

Although we do not know the exact origin of the effective noise observed here and in previous work \cite{Medford2013a, Medford2013}, we note that the overall noise levels need to be reduced by several orders of magnitude to allow high-fidelity entangling gates~\cite{Srinivasa2016}.
As a cautionary advice against the overuse of partial sweet spots, we note that for any qubit tuned smoothly by $N$ (in our work five) gate voltages one can always (i.e., for any operating point) define at least $N-1$ (in our work 4) independent control parameters that to first order do not influence the qubit splitting. This underlines the importance of careful analysis of noise sources and noise correlations~\cite{Szankowski2016} in determining optimal working points of qubits~\cite{Cywinski2014}.

For the 530~MHz tuning the SRX qubit appears to outperform the RX qubit for $n>8$, indicating that the spectral noise density at higher frequencies, filtered by the CMPG sequence~\cite{Martinis2003,Cywinski2008,Soare2014,Malinowski2017}, may indeed be reduced for the SRX qubit.
The scaling of $T_2^\mathrm{CPMG}$ with (even) number of pulses appears to follow a power law. Although the exponent (0.77$\pm$0.07) for the SRX data is consistent with values reported for RX qubits~\cite{Medford2013a}, a spectral interpretation may need to take into account unconventional decoherence processes that can occur at sweet spots, such as non-Gaussian noise arising from quadratic coupling to Gaussian distributed noise and the appearance of linear coupling to noise arising from low-frequency fluctuations around a sweet spot~\cite{Bergli2006,Cywinski2014}.

In conclusion, we have operated a triple-dot resonant exchange qubit in a highly symmetric configuration.
At the three-dimensional sweet spot the overall sensitivity of the qubit frequency to five control voltages is reduced by an order of magnitude, but resonant operation of the qubit is technically more demanding. 
For weak resonant driving the quality of Rabi oscillations show no significant improvement due to the dominant contributions of nuclear Overhauser gradients to fluctuations of the qubit splitting, motivating the future use of nuclear-spin-free semiconductors. 
For strongly driven Rabi oscillations and CPMG decoupling sequences the coherence times are significantly shorter than expected from instrumentation noise alone and Overhauser fluctuations, suggesting that recent theory must be extended to include the dependence of drive strength on control voltages. 
An optimization of gate lever arms and materials' charge noise may then allow non-resonant operation of multi-qubit structures that take advantage of highly symmetric configurations of triple-dot qubits.

This work was supported by the Army Research Office, the Villum Foundation, the Innovation Fund Denmark, and the Danish National Research Foundation.


\end{document}